\documentclass[12pt,psfig]{iopart}   %%% IOP journal style  %%%
\usepackage{iopams}

\usepackage{color} 
\usepackage{graphicx}

\usepackage{psfig}
\usepackage{pstricks}

\newcommand{\beq}{\begin{equation}}
\newcommand{\eeq}{\end{equation}}
\newcommand{\be}[1]{\begin{equation}\label{#1}}
\newcommand{\ee}{\end{equation}}
\newcommand{\continue}{\nonumber \\ }
\newcommand{\bea}{\begin{eqnarray}}

\newcommand{\eea}{\end{eqnarray}}

\newcommand{\bm}[1]{\mbox{\boldmath{$ #1$}}}
\newcommand{\MatrixIII}[9]{
   \pmatrix{ {#1}  &  {#2} &  {#3} \cr
             {#4}  &  {#5} &  {#6} \cr
             {#7}  &  {#8} &  {#9} \cr} }

\newcommand{\refeq}  [1] {(\ref{#1})}

\newcommand{\reffig} [1] {fig.~\ref{#1}}

\newcommand{\refsect}[1] {sect.~\ref{#1}}

\newcommand{\refSect}[1] {Sect.~\ref{#1}}

\newcommand{\ed}{\mathbf{d}}
\newcommand{\field}{\mathbf{U}}
\newcommand{\memField}{\mathbf{u}}

\newcommand{\volel}{\mathbf{ dV}}
\newcommand{\hdual}[1]{\mathbf{dS}_{#1}}
\newcommand{\oneform}[2]{{\bomega^{#1}}_{ #2}}
\newcommand{\bas}[1]{\bTheta^{#1}}
\newcommand{\stressForm}[1]{\mathbf{f}^{#1}}
\newcommand{\surfGrad}{\widetilde{\bnabla}}
\newcommand{\midbas}[1]{\bphi^{#1}}
\newcommand{\midoneform}[2]{{\widetilde{\bomega}^{#1}}_{ #2}}
\newcommand{\lieDer}{\mathcal{L}}
\newcommand{\dmod}{\check{d}}
\begin{document} 
\title{Moving frames applied to shell elasticity}

%\today 
%\centerline{and}
\vspace{-5mm}
\author {Niels S\o ndergaard}
\address{Div. of Mathematical Physics,\\
LTH , Sweden.}
\date{\today}
\ead{niels.sondergaard@matfys.lth.se}

\begin{abstract}
Exterior calculus and moving frames are used to describe curved elastic shells. The kinematics follow from the Lie-derivative on forms whereas the dynamics  via  stress-forms. 
\end{abstract}

% \tableofcontents

%\begin{minipage}{0.5\textwidth}

\section{Introduction}
\label{sec:introduction}
Shell theory  reduces the three-dimensional elasticity of a shell to an effective two-dimensional theory of its middle-section. This article presents a new short derivation of the shell equations  of Kirchoff-Love based on the method of moving frames and exterior calculus.

Early shell theory  dealt with church bells. Lord Rayleigh studied vibrating shells motivated by the question whether or not a full musical scale could be played on bells \cite{RaylBell}. Vibrations of shells for  airplanes and cars continue to be important in civil engineering.  There, the numerical  modelling of shells remain an active area of research relevant for e.g. car crashes. Also computer scientists have joined the  efforts in order to present realistic computer graphics \cite{discreteShell}. Lately,  shell theory has been applied to cell membranes \cite{hairCell} and even nanotubes \cite{yakobson}. 

There are several approaches to shell theory.  Kraus  presents a  variational method in  \cite{kraus}.  Equivalently, assuming Newton's second law for an infinitesimal volume element and integrating over  the thickness of the shell lead to the same results,  see e.g. \cite{fluegge}.   Unfortunately the calculations are complicated: the dynamic part of this well-known law involves a covariant derivative of the stress tensor and requires intricate manipulations of   Christoffel symbols for the full shell, respectively its middle-section.

However, a general observation is that complexities in tensor calculus can be avoided if it is possible to work in an orthonormal frame, i.e. in  the case of Riemannian manifolds. Then the number of connection coefficients are drastically reduced and calculations become simpler and more transparent.  For example,   Maxwell's equations in  electrodynamics and calculations of curvature in general relativity become much simpler when using the exterior calculus of differential forms and Cartan's method of moving frames \cite{grav}.   It is in this spirit we shall discuss shell theory.

\subsection{Thin shell assumptions  }
\label{sec:stressAndstrain}
For simplicity, we follow the  assumptions of  Kirchoff and Love for a thin shell: vertical coordinate $z$ and intrinsic coordinates $\alpha_1,\alpha_2$:
\begin{enumerate}
\item thickness small compared to radii of curvature
\item small displacements
\item vanishing normal stress $\sigma_{33} =0$
\item preservation of normals $\epsilon_{i3} =0$
\item linear dependence of membrane field $U^{i} = u^{i}(\alpha^{1},\alpha^{2},t) + z \beta^{i}(\alpha^{1},\alpha^{2},t) $
\item constant dependence of flexural field $W = w(\alpha^{1},\alpha^{2},t) $
\end{enumerate}

\subsection{Outline}
The presentation centers around a proof of the shell equations. Some familiarity with differential forms and continuum mechanics is assumed. \refSect{sec:geom} introduces the  geometry  via differential forms, \refsect{sec:strain} treats the kinematics of shell deformation   and \refsect {sec:elasto} the equations of motion.  A summary and some concluding
remarks will be given in the last section. 

\section{Geometry}
\label{sec:geom}

\subsection{Geometry of the mid-section $\mathcal{M}$}

The shell equations must be covariant and therefore it is enough to validate them in one set of coordinates. Hence,  on the mid-section $\mathcal{M}$ we  can choose    {\it lines of curvature}  coordinates $\alpha^1$ and $\alpha^2$ having  axes aligned with the principal directions of curvature.  Such coordinates  can always be chosen locally on a two-dimensional surface \cite{curvCoord}. When the radii of curvature differ $R_1 \neq R_2$ the axes can be chosen uniquely. Otherwise, at an umbilic 
point $R_1 = R_2 \neq \infty$ or at a planar point $R_1 = R_2 = \infty$, we shall assume a choice made.  Write in these coordinates, the {\it mid-section metric} in the Lam\'{e}-form
\beq
ds^2 = A_1^2 \, (d\alpha^1)^2 + A_2^2 \, (d\alpha^2)^2 \,.
\eeq
The $A_i$ are called Lam\'{e}-parameters. 
\begin{figure}[htbp]
   \centering
   \includegraphics[height=8cm]{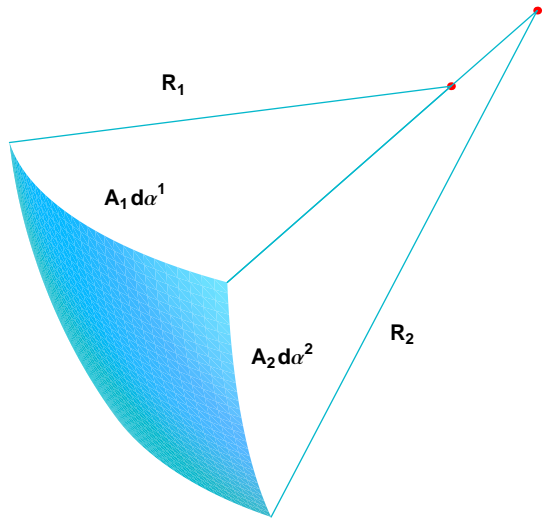} % requires the graphicx package
   \caption{Middle section segment with curvature radii $R_1,R_2$ and increments  $A_1 d\alpha^1,A_2 d\alpha^2$.}
   \label{fig:shellElement}
\end{figure}
Instead of working directly with the coordinates $\alpha^1$ and $\alpha^2$ as done traditionally  we shall work with a frame derived from these particular coordinates.
Thus define an orthonormal frame on the mid-section via the forms 
\beq
\midbas{a}= A_{a} d\alpha^{a} \,.
\eeq
Then the metric of the mid-section  {\it the first fundamental tensor} takes the simple form
\beq
\mathbf{a} = g_{a b} \, dx^a \otimes dx^b \equiv \midbas{1}\otimes\midbas{1}+ \midbas{2} \otimes \midbas{2} \equiv \delta_{a b} \, \midbas{a} \otimes \midbas{b}  \,
\eeq
and the curvature tensor of the mid-section  {\it the second fundamental tensor} becomes
\beq
\mathbf{d} = d_{a b} \, dx^a \otimes dx^b \equiv  \frac{1}{R_1}\,\midbas{1}\otimes\midbas{1} +\frac{1}{R_2}\, \midbas{2} \otimes \midbas{2}  \,.
\eeq

\subsection{Geometry of the shell $\mathcal{S}$}
Denote the  coordinates on the {\it three-dimensional shell} $\mathcal{S}$:  $\alpha^{1}, \alpha^{2}$ and {z}. The latter coordinate is  perpendicular to the middle section of the shell, see \reffig{fig:shellElement}. 
The direction of the $z$-axis follows that of \cite{kraus} but is opposite from e.g. \cite{fluegge,niordsson}.  
The thickness of the shell is called $h$.  $z$ equal $\pm h/2$  defines the upper respective  the lower section of the shell. Likewise $z=0$ corresponds to the mid-section $\cal{M}$. 
The {\it shell  metric} is 
\beq
\label{eq:metric}
ds^{2} = A_{1}^{2} (1+z/R_{1})^{2} (d\alpha^{1})^{2} + A_{2}^{2} (1+z/R_{2})^{2} (d\alpha^{2})^{2} + dz^{2} \, .
\eeq
The factors $1+z/R_{i}$ are understood as follows: An increment on the middle surface in the direction $i$ spans the angle
\[
d\varphi^{i} = A_{i} d\alpha^{i} /R_{i}
\]
so the increment at the elevated position is 
\[
dx^{i}(z) = (R_{i}+ z) \, d\varphi^{i} = A_{i} (1+ z/R_{i}) d\alpha^{i} \, .
\]
Finally, we shall assume furthermore the  existence of derivatives of $A_i, R_j$.

\subsubsection{Orthornormal frame of $\mathcal{S}$}
We introduce a three-dimensional frame as done for  the mid-section  instead of working with the coordinates $\alpha^1,\alpha^2$ and $z$ as in  \cite{kraus,goldenvaiser} .
Thus, from \refeq{eq:metric} define  an orthonormal basis of one-forms as:
\beq \label{eq:OneFormDef}
\fl \bas{1} = A_{1}\, (1+z/R_{1})\, d\alpha^{1} \, ,  \qquad \bas{2} = A_{2}\,(1+z/R_{2})\, d\alpha^{2} \qquad \mbox{and} \qquad \bas{3}=dz \, .
\eeq
From the one-forms construct the volume element
\beq
\volel = \bas{1} \wedge \bas{2} \wedge \bas{3} = A_{1} A_{2} \, (1+z/R_{1})(1+z/R_{2})\, d\alpha^{1} \wedge d\alpha^{2}\wedge dz
\eeq
The Hodge-dual forms to the one-forms are oriented area-elements 
\beq
\hdual{a} = \star \bas{a}
\eeq
and fulfill
\beq
\label{eq:dualForm}
\bas{a} \wedge \hdual{b} = \delta^{a}_{b} \, \volel \, .
\eeq
In component form
\beq
\hdual{i}=  \varepsilon_{i j k} \bas{j} \otimes \bas{k} 
\eeq
with $\varepsilon_{i j k}$ the totally antisymmetric tensor. Note, in the following we  reserve $\epsilon_{i j}$ for another tensor, the strain, and hopefully no confusion will arise.  As the area-forms are not closed  $d(\hdual{i}) \neq 0$ they are not exact  $\hdual{i} \neq d(\mathrm{1-form})$, and hence  the notation of these area forms is slightly misleading although  conventional. 

\subsubsection{The connection of  $\mathcal{S}$}
The one-forms of the frame lead to the connection coefficients via the condition of vanishing torsion  \cite{grav}:
\beq \label{eq:TorsCond}
d\bas{i} + \oneform{i}{j} \wedge \bas{j} = 0 \,.
\eeq
Calculating $d\bas{i}$ from \refeq{eq:OneFormDef}  the connection form $\oneform{i}{j}$  is found by inspection: 
\bea \label{eq:connection}
\fl \bomega = (\oneform{i}{j}) \\ \continue \nonumber
\fl = \MatrixIII{0}{ \frac{(A_{1}(1+z/R_{1}))_{,2} \bas{1}-(A_{2} (1+z/R_{2}))_{,1} \bas{2} }{A_{1} A_{2}(1+z/R_{1})(1+z/R_{2})} }{\frac{\bas{1}}{R_{1}+z} }{ \frac{-(A_{1}(1+z/R_{1}))_{,2} \bas{1}+(A_{2} (1+z/R_{2}))_{,1} \bas{2} }{A_{1} A_{2}(1+z/R_{1})(1+z/R_{2})} }{0}{\frac{\bas{2}}{R_{2}+z} }{-\frac{\bas{1}}{R_{1}+z} }{-\frac{\bas{2}}{R_{2}+z} }{0} \\ \continue \nonumber
\fl =   \MatrixIII{0}{ \frac{A_{1,2} \bas{1}/(1+z/R_{1})-A_{2,1} \bas{2}/(1+z/R_{2}) }{A_{1} A_{2}} }{\frac{\bas{1}}{R_{1}+z} }{ \frac{-A_{1,2} \bas{1}/(1+z/R_{1})+A_{2,1} \bas{2}/(1+z/R_{2}) }{A_{1} A_{2}} }{0}{\frac{\bas{2}}{R_{2}+z} }{-\frac{\bas{1}}{R_{1}+z} }{-\frac{\bas{2}}{R_{2}+z} }{0} \, ,
\eea
where derivatives are denoted with a comma  $\partial f / \partial \alpha^{i} =f_{,i} $. The final simplification in \refeq{eq:connection}  follows from  the Gauss-Codazzi equations,  shortly discussed in the following. 

Finally, note  that in the subsequent calculation of various covariant derivatives the number of connection coefficients $\Gamma_{k j}^i \equiv  \iota_{\mathbf{e}_k}    \oneform{i}{j} $ are far less in an orthonormal  frame as compared to a general coordinate basis as these coefficients  for one pair of indicices now are antisymmetric and not just symmetric.

\subsubsection{Gauss-Codazzi}
 There exist constraints on the functions $A_{i}$ and $R_{i}$ in the metric \refeq{eq:metric}  for defining a valid surface. These equations come about by expressing the curvature via the connection. Even though the shell is curved the metric \refeq{eq:metric} is just a metric for the flat three dimensional space ${\mathbb{ R}}^{3}$.  Consequently the curvature is zero:
\beq
0={\bOmega^{i}}_{j} = d\oneform{i}{j} + \oneform{i}{k} \wedge \oneform{k}{j}
\eeq
For instance for $i=1$ and $j=3$ 
\beq d\oneform{1}{3} =(A_{1}/R_{1})_{,2} \, d\alpha^{2} \wedge d\alpha^{1} \eeq
and
\beq \oneform{1}{2} \wedge \oneform{2}{3} =\frac{(A_{1} (1+z/R_{1}))_{,2} }{R_{2} +z} \, d\alpha^{1} \wedge d\alpha^{2} \, , \eeq
so 
\beq 
\label{eq:CodazziShell0}
\frac{\left( A_{1} (1+z/R_{1})\right)_{,2}}{ R_{2}} = \left(\frac{A_{1}}{R_{1}}\right)_{,2} \, (1+z/R_{2}) \, . \eeq
At the middle surface $z=0$ the classical Codazzi equation is found
\beq
\label{eq:Codazzi}
\frac{A_{1,2}}{R_{2}} =\left(\frac{A_{1}}{R_{1}}\right)_{,2} \, .
\eeq
Applying \refeq{eq:Codazzi} to \refeq{eq:CodazziShell0} gives
\beq
(A_{1} (1+z/R_{1}))_{,2} = A_{1,2} \, (1+z/R_{2}) \,
\eeq
used in \refeq{eq:connection}.
Likewise other identities between $A_{1},A_{2},R_{1}$ and $R_{2}$ hold generalizing the Gauss-Codazzi-equations to the shell metric. By considering $i=1$ and $j=2$ the result of Gauss follows:
\beq \label{eq:IntrinsicEmbed}
\left(\frac{A_{1,2}}{A_{2}}\right)_{,2} +\left(\frac{A_{2,1}}{A_{1}}\right)_{,1} =-\frac{A_{1} A_{2}}{R_{1 } R_{2}} \, .
\eeq  
Hence using \refeq{eq:IntrinsicEmbed}, the total curvature $K \equiv \frac{1}{R_{1}R_{2}}$ related to the embedding may be expressed entirely in terms of the intrinsic functions $A_{i}$ belonging to  metric of the surface. This is the lines of curvature coordinates version of  Gauss's theorema egregium \cite{curvCoord}.

\subsubsection{Covariant Derivative}
Having defined the connection $\oneform{}{}$,  the covariant derivative acting on forms is
\beq
\bnabla = \ed + \bomega \, ,
\eeq
where $\ed$ is the flat exterior derivative and $\bomega$ acts from the left with the wedge-product. 
From the condition of no torsion \refeq{eq:TorsCond} follows that the covariant derivative vanish on frame co-vectors respective tangent-vectors. The latter holds since the  metric is  covariantly constant.  Thus
\beq \label{eq:ConstBas}
\bnabla \bas{i} = 0 \, .
\eeq
To exploit this structure we shall  consequently define  quantities of interest as forms.

\subsubsection{The derived connection of  $\mathcal{M}$}
The mid-section $\mathcal{M}$  is embedded in the shell $\mathcal{S}$:
\beq
i: \mathcal{M} \rightarrow \mathcal{S}
\eeq 
so restricting  differential forms to the mid-section is the {\it pull-back}
\beq
i^*: \Lambda^*( \mathcal{S}) \rightarrow \Lambda^*(\mathcal{M}) \,.
\eeq
This restriction is obtained by putting $z=0$. For instance
\beq
i^*(\bas{a}) = \midbas{a} \qquad \mathrm{and} \qquad i^*\bas{3} =0 \,.
\eeq
Likewise, pulling back  the connection of the three-dimensional shell $\oneform{i}{j}$ to the mid-section $i^*$ gives {\it the} two-dimensional connection $\midoneform{a}{b}$ of $\mathcal{M}$.
Namely
\beq
\label{eq:NoTorsionMid}
d \midbas{a} = -\oneform{a}{b}|_{z=0} \wedge \midbas{b}  \equiv - \midoneform{a}{b} \wedge \midbas{b}
\eeq
with $a,b = 1,2$ only, where the "tilde" shall be used in the following when referring to the mid-section.  Formally  \refeq{eq:NoTorsionMid} holds since the exterior derivative commutes with the pull-back 
\beq 
d \, i^* = i^* \,  d
\eeq
and
\beq
d \midbas{a} = d(i^*\bas{a}) = i^* d\bas{a} = i^*(-\oneform{a}{i} \wedge \bas{i}) = - i^*(\oneform{a}{b}) \wedge \midbas{b} \, .
\eeq 
Thus, all connection coefficients of the mid-section $\widetilde{\Gamma}^i_{j k}$ are contained in the single one-form 
\beq \label{eq:connectionMid}
\midoneform{1}{2} = \frac{A_{1,2} \midbas{1} - A_{2,1} \midbas{2}}{A_1 A_2} \,.
\eeq
where
\beq
\midoneform{a}{b} \equiv \widetilde{\Gamma}^a_{c b} \, \midbas{c} = -\midoneform{b}{a} \,.
\eeq
Finally, these connection coefficients allow us to define the covariant derivative on the mid-section, $\surfGrad$. 

\section{Shell kinematics} 
\label{sec:strain}

This section describes the kinematics  of a shell. The appropriate deformation field is the strain field  $\bepsilon$. This field corresponds to relative displacements or {\bf e}xtensions. 

\subsection{The displacement field $\field$}
The displacement is a tangent vector field and hence contra-variant. According to Kirchoff-Love:
\beq \label{fieldDef}
\fl \field = U^{1} \mathbf{e}_{1} + U^{2} \mathbf{e}_{2}+ W \mathbf{e}_{3}\equiv(u^{1} +z \beta^{1}) \mathbf{e}_{1} + (u^{2}+z \beta^{2}) \mathbf{e}_{2}+ w \mathbf{e}_{3}
\eeq
More general functional forms exist: e.g. quadratic dependence in the flexural field $w$ and even  general Taylor expansions in $z$ have been proposed by Vlasov \cite{kraus}.

\subsubsection{Rotation angles $\beta^{i}$}
If  the $\bbeta$ in \refeq{fieldDef} is sufficiently small its components play the role  of  angles. Thus when seen from the middle section the displacement at height $z$ in the direction $i$ equals $z \, d\varphi^{i} \approx z \beta^{i}$. 
However, in Kirchoff-Love's theory,  the rotations $\beta^{i}$ are not considered as independent fields but can be found from $u^i$ and the gradient of $w$.  This we show in the following \refsect{sec:strainExpand} using  the assumption of vanishing normal strain $\epsilon_{i3}=0$.

\subsection{Strain field: connection to the Lie-derivative}
\label{sec:strainAndLie}
In linear elasticity the strain field follows from the displacement field by an application of a Lie-derivative on the metric \cite{frankel}
\beq \fl
\bepsilon = \epsilon_{i j}\, \bas{i} \otimes \bas{j} = \frac{1}{2} \, \mathcal{L}_{\field} ( \mathbf{g}) \qquad \mbox{with} \qquad \mathbf{g} \equiv g_{i j} dx^i \otimes dx^j= \delta_{j k} \bas{j}\otimes\bas{k}
\eeq
The Lie-derivative $\mathcal{L}$  is with respect to the physical displacement field $\field$. 

The strain field is calculated with respect to the orthonormal frame \refeq{eq:OneFormDef}. 
As $\mathcal{L}$ is a derivation:
\beq
\mathcal{L} (\bas{j} \otimes \bas{k}) = \mathcal{L} (\bas{j}) \otimes \bas{k} + \bas{j} \otimes   \mathcal{L}( \bas{k}) \,
\eeq
it suffices to investigate its action on  a co-vector $\bas{j}$.
Using Cartan's formula \cite{frankel} for the Lie-derivative and \refeq{eq:ConstBas}:  
\bea
\fl  \mathcal{L}_{\field}(\bas{j}) & \equiv (d \, \iota_{\field} + \iota_{\field} \, d) \bas{j} = d U^{j}+\iota_{\field} (d \bas{j}) = d U^{j} - \iota_{\field}(\oneform{j}{k} \wedge \bas{k}) \continue
      &= d U^{j}+ U^{k} \oneform{j}{k} - (\iota_{\field} \oneform{j}{k}) \bas{k} = \bnabla U^{j} - (\iota_{\field} \oneform{j}{k}) \bas{k} \, .
\eea
The antisymmetry $\oneform{j}{k}=-\oneform{k}{j}$ gives the familiar result from continuum mechanics:
\beq \label{eq:StrainExpr} \fl
\bepsilon = \frac{1}{2} (\bnabla U^{j} \otimes \bas{k} + \bas{k}\otimes \bnabla U^{j}) g_{j k} = \frac{1}{2} (\bnabla \otimes \bas{j} {U}_{j} + {U}_{j} \bas{j}\otimes \bnabla ) = \frac{1}{2} (\bnabla \field + \field \bnabla) \, .
\eeq
In \refeq{eq:StrainExpr} the metric lowers the indices of the displacement field coordinates from a tangent vector to a co-vector. Sometimes this is made explicit  by writing $\field=\field^\flat$.

\subsection{Lie-derivative gives variation of curvature}
\label{sec:strainExpand}
 We wish to emphasize the connection of the strain in shells  to the variation of the second fundamental tensor. This is done using the Lie-derivative instead of the conventional expression  of the strain \refeq{eq:StrainExpr}.

\paragraph{Expanding the shell metric} 
To find the strain with respect to the mid-section the full shell metric \refeq{eq:metric} is written out in terms of simpler mid-section one-forms 
\bea
\fl \mathbf{g} &= \midbas{1}\otimes\midbas{1} + \midbas{2}\otimes\midbas{2} + \bas{3}\otimes\bas{3}  \continue
\fl  &+ 2\,z  \, \left( \frac{1}{R_1} \,  \midbas{1} \otimes \midbas{1} +   \frac{1}{R_2} \,  \midbas{2} \otimes \midbas{2}\right) +  z^2 \,  \left( \frac{1}{R_1^2} \,  \midbas{1} \otimes \midbas{1} +   \frac{1}{R_2^2} \,  \midbas{2} \otimes \midbas{2}\right) \continue
& \equiv \mathbf{a} +  \bas{3}\otimes\bas{3}   + 2 z \, \mathbf{d} + z^2 \, \mathbf{d}^2 \,
\eea
with $d^2_{a b} \equiv d_{a c} \, d^{c}_b$.
Deriving with  the Lie-derivative gives
\bea
\label{midsectExpansion}
\fl \bepsilon &\equiv \frac{1}{2}\,\lieDer_{\field}(\mathbf{g}) \continue
& = \frac{1}{2}\,\lieDer_{\field}(\mathbf{a}) +\frac{1}{2}\, (\bnabla w \otimes \bas{3} + \bas{3} \otimes \bnabla w)+ \lieDer_\field (z \, \mathbf{d})+ \frac{1}{2} \, \lieDer_\field(z^2 \, \mathbf{d}^2) \continue
&= \frac{1}{2}\,\lieDer_{\field}(\mathbf{a}) + w \mathbf{d} + z \, \lieDer_\field ( \mathbf{d}) + z \,  w \,    \mathbf{d}^2 +\frac{1}{2}\, (\bnabla w \otimes \bas{3} + \bas{3} \otimes \bnabla w)+O(z^2) \continue
 \eea
Above, the Lie-derivative contracts  the mid-section forms $\midbas{a}$ via the interior derivative $\iota_\field$. Hence,   it is advantageous to express the frame coordinates of the displacement field with respect to the mid-section:
 \bea \fl
( u^i + z \, \beta^i) e_i &\equiv ( u^i + z \, \beta^i) \frac{1}{(1+ z/R_i) A_i} \frac{\partial}{\partial \alpha^i} \continue & = ( u^i + z \, \beta^i)(\delta^j_i-z \, d^j_i)  \, \frac{1}{A_j} \frac{\partial}{\partial \alpha^j} + O(z^2) \continue
&\equiv (u^i + z(\beta^i - d^i_j \, u^j) ) \, \tilde{e}_i  +O(z^2) \equiv (\widetilde{u}^i + z \widetilde{\beta}^i) \tilde{e}_i + O(z^2)
 \eea
 with
 \bea
 \widetilde{\mathbf{u}} &= \mathbf{u} \\
 \widetilde{\bbeta} &= \bbeta- \mathbf{d} \cdot \memField \,.
 \eea
 To proceed consider the first term in \refeq{midsectExpansion} 
 \beq \label{Eq:midStrain}
 \frac{1}{2}\,\lieDer_{\field}(\mathbf{a}) = \frac{1}{2}\,\lieDer_{\memField+ z \bbeta}(\mathbf{a}) = \frac{1}{2}\,\lieDer_{\memField}(\mathbf{a}) +\frac{1}{2}\,\lieDer_{ z \bbeta}(\mathbf{a}) 
 \eeq
The first term in \refeq{Eq:midStrain} reduces to a membrane strain expressed in mid-section coordinates
\beq
\label{eq:memStrain}
\frac{1}{2}\,\lieDer_{\field}(\mathbf{a}) = \frac{1}{2} (\widetilde{u}_{a;b} + \widetilde{u}_{b;a})\, \midbas{a} \otimes \midbas{b} = \frac{1}{2} (\surfGrad \otimes \memField  + \memField \otimes \surfGrad)
\eeq
with the corresponding covariant derivative.
The second term in \refeq{Eq:midStrain} gives a normal strain plus a  strain similar to \refeq{eq:memStrain} of order $O(z)$:
\bea \label{eq:rotLieDer}
\frac{1}{2}\,\lieDer_{ z \bbeta}(\mathbf{a}) =\frac{1}{2}\, \widetilde{\beta}^a (dz \otimes \midbas{b} + \midbas{b} \otimes  dz)\, \delta_{a b}+z \, \frac{1}{2} \, (\surfGrad \otimes \widetilde{\bbeta}  + \widetilde{\bbeta} \otimes \surfGrad) \continue
 =\frac{1}{2}\,  (\bas{3} \otimes \widetilde{\bbeta} + \widetilde{\bbeta} \otimes  \bas{3})\, +z \, \frac{1}{2} \, (\surfGrad \otimes \widetilde{\bbeta}  + \widetilde{\bbeta} \otimes \surfGrad)
\eea
As  $\frac{\partial w}{\partial z}=0$ the normal strain $\epsilon_{3 3}$ vanishes in \refeq{midsectExpansion}. So effectively
\beq
\bnabla{w} = dw = dw(\alpha^1,\alpha^2)  = \surfGrad w \,.
\eeq
Then 
the requirement of  {\it vanishing normal shear strain}  $\epsilon_{a 3}=0$  with \refeq{midsectExpansion} and \refeq{eq:rotLieDer} reduces to
\beq
\widetilde{\bbeta} = -\surfGrad w +O(z) \,.
\eeq 
An $O(z)$-correction is indicated, since $\mathcal{L}_{z \bbeta} (\mathbf{d})$ from $\mathcal{L}_{\field} (\mathbf{d})$ produces such a term.
In conclusion 
\bea \label{eq:strainExpansion}
\fl \bepsilon & \equiv \bepsilon^{(0)} + z\, \bepsilon^{(1)}=\frac{1}{2} (\surfGrad \otimes \memField  + \memField \otimes \surfGrad ) + w \mathbf{d} \continue & -  z \left( \frac{1}{2}(\surfGrad \otimes \surfGrad w + \surfGrad w \otimes \surfGrad) - \mathbf{d}^2 w - \lieDer_\memField(\mathbf{d}) \right) + O(z^2) \continue
& = \frac{1}{2} (\surfGrad \otimes \memField  + \memField \otimes \surfGrad ) + w \mathbf{d}  \continue
&  -  z \left( \surfGrad \otimes \surfGrad w - \mathbf{d}^2 w - \lieDer_\memField(\mathbf{d}) \right) + O(z^2)\, .
\eea
This result agrees with that in \cite{niordsson}  ( where $\mathbf{d} := - \mathbf{d} $ ) when the Lie-derivative in                                 \refeq{eq:strainExpansion}  is written out in full.  Using \cite{hawking} or performing  similar calculations as in \refsect{sec:strainAndLie}  gives the standard result 
\beq
\lieDer_\memField(\mathbf{d}) = \left( d_{a b;c} u^c + d_{a }^{c} u_{b;c} + d^{c}_{ b} u_{a;c} \right) \midbas{a}\otimes \midbas{b} \,
\eeq
with the covariant derivative corresponding to the mid-section.
This Lie-derivative now measures the variation of the second fundamental tensor, the curvature tensor,  with respect to membrane fields.  It represents the change of curvature when only stretching.

\paragraph{First versus second fundamental tensor}
To summarize, the Lie-derivative of the first fundamental tensor $a_{a b}$ produces  parts of the leading in-plane strain $\bepsilon^{(0)}$, whereas on  the second fundamental tensor $d_{a b}$   parts of the bending tensor  $\bepsilon^{(1)}$.  \cite{fluegge,niordsson} introduce various measures of variation of the second fundamental tensor and   linearize these. The same results are obtained here formally, where  the Lie-derivative  eventually acts on the curvature tensor in the expansion of the metric. 

\paragraph{How unique is the bending tensor? }
The discussion in this text considers the variation of $\mathbf{d}=d_{a b} \midbas{a}\otimes \midbas{b}$ under deformation.
Since  the Lie-derivative does not commute with raising and lowering indices, in fact $\mathcal{L}_{\field}( \mathbf{g}) = 2 \bepsilon$, the various measurements of bending defined from the variation of $d_{a b}, d_{a}^{b}$ or $d^{a b}$  differ with terms of the form $\bepsilon \cdot \mathbf{d}$ \cite{niordsson}.  
For instance \cite{fluegge} considers  the variation of  $ d^c_b$.

\paragraph{What dominates: bending or stretching ?}
The most important terms in the strains \refeq{eq:strainExpansion} at short wave lengths, the {\it principal part}, can be identified as the highest derivatives. The term of the  strain proportional to the vertical displacement $ \bepsilon^{(1)}$  has as  its principal term  $ \surfGrad \otimes \surfGrad w  $    associated with bending and arises from the $z \bbeta$ contribution to the \hbox{in-plane} strain. On the other hand, the principal part of $\bepsilon^{(0)}$  comes from  the tangential stretching field described by the $\surfGrad \odot \field $, the  in-plane strain.

\section{Shell dynamics}
\label{sec:elasto}
In this section the stresses and moments in the shell are found. We follow  \cite{fluegge} but use exterior calculus instead.

\subsection{Force conservation in the bulk} 
\label{derive}
Forces acting on general surface elements $\mathbf{dS}$
are described in continuum mechanics \cite{lAndl,auld} in terms of the  stress tensor $\bsigma$.
Thus the infinitesimal force on a surface element equals
\beq
\label{def:stressContinuum}
\mathbf{df }=\mathbf{dS} \cdot \bsigma \,.
\eeq
In the bulk, the force per unit volume
becomes
\beq
\label{microEquilStandard}
\bnabla \cdot \bm{\sigma}({\field}) + {\bf X}= \rho \, \frac{\partial^2}{\partial t^2}{\field} \, ,
\eeq
where ${\bf X}$ is an external volume body force. In this section we focus on the left-hand side of \refeq{microEquilStandard}, that is the {\it dynamic} part of the equations of motion.

In exterior calculus it is convenient to collect  the forces on an infinitesimal element  via the {\it stress-form } \cite{frankel}
\beq
\label{def:Stress}
\stressForm{i}=(\hdual{} \cdot \bsigma)^{i} = \sigma^{1 i} \hdual{1} + \sigma^{2 i} \hdual{2} + \sigma^{3 i} \hdual{3} \, .
\eeq
This form is a two-form and when integrated over a surface it gives the  corresponding force (here in direction $i$, which by convention is put as the last index in the $\sigma^{i a}$). In continuum mechanics  $\stressForm{i}$ is infinitesimal as in \refeq{def:stressContinuum} but in exterior calculus the notation does not indicate so as this form is not exact in general.   

The total force is  written as
\beq
\mathbf{f} = \stressForm{i} \, \mathbf{e}_i \equiv \stressForm{i} \otimes \mathbf{e}_i
\eeq
Its covariant derivative is  simple
\beq
\label{microEquil0}
\bnabla \mathbf{f} = \bnabla \stressForm{i} \otimes \mathbf{e}_i + \stressForm{i} \otimes \bnabla  \mathbf{e}_i = \bnabla \stressForm{i} \otimes \mathbf{e}_i
\eeq
as the frame and its dual are constant and similar for $\mathbf{X}, \mathbf{\ddot{u}}$.

In differential form \refeq{microEquilStandard} reads for the component $i$
\beq
\label{microEquil}
\bnabla \stressForm{i} + {\bf X}^{i} \volel = \rho \ddot{u}^{i} \volel \,.
\eeq
\subsection{Covariant derivative of stress form}
\label{sec:covDerStressForm}
First we discuss tangential directions. Hence, make a split in tangential and normal (indices $a,b \in \{ 1, 2\}$ versus $3$) in \refeq{microEquil}:
\bea \label{eq:nablaStressForm}
\fl \bnabla \stressForm{a} = \bnabla (\sigma^{1 a} \hdual{1} + \sigma^{2 a} \hdual{2} + \sigma^{3 a} \hdual{3}) \\  \nonumber
\fl = \partial_{1}(\sigma^{1 a} A_{2}(1+z/R_{2})) d\alpha^{1}\wedge d\alpha^{2} \wedge dz +\partial_{2}(\sigma^{2 a} A_{1}(1+z/R_{1})) d\alpha^{2}\wedge dz \wedge d\alpha^{1}    \\  \nonumber
\fl  + \partial_{3}(\sigma^{3 a}(1+z/R_{1})(1+z/R_{2})A_{1} A_{2}) dz \wedge d\alpha^{1} \wedge d\alpha^{2}+ \oneform{a}{b} \wedge \stressForm{b} + \oneform{a}{3} \wedge \stressForm{3}   
\eea
Consider  the first component $a=1$. A simplification occurs in the last two terms of \refeq{eq:nablaStressForm}, since e.g. 
\hbox{$\oneform{a}{3} =  \bas{a}/(R_{a}+z)$} restricts the relevant term in $\stressForm{3}$ to $\sigma^{a 3} \hdual{a}$   by \refeq{eq:dualForm}. Likewise the second last term has $b=2$, where
\beq
\label{eq:oneform12}
\oneform{1}{2} = \frac{1}{A_{1} A_{2}} \left(A_{1,2} \bas{1}/(1+z/R_{1})-A_{2,1} \bas{2}/(1+z/R_{2}) \right)
\eeq 
picks out corresponding terms of $\stressForm{2}$, that is $\sigma^{2 1}$ for the first term in \refeq{eq:oneform12} and $\sigma^{2 2}$ for the second. 

\subsection{Momentum conservation}
\label{redDim}
The degree of freedom in the direction perpendicular to the shell  is eliminated by integrating \refeq{microEquil}  over the thickness coordinate,  $\int_{z=-h/2}^{h/2}$.

\subsubsection{Integrated stress tensor}
Consider the force per length with respect to an edge along the middle section. For example, assume the edge is perpendicular to the first direction. That force density equals,
\beq
N^{1 a} \equiv \frac{\int_{-h/2}^{h/2} \rmd S_{1}(z) \, \sigma^{1 a} }{A_{2} \, \rmd \alpha^{2}} =\int_{-h/2}^{h/2} \rmd z \, \sigma^{1 a} \, (1+z/R_{2})\, .
\eeq
The notation is such that the first index denotes the direction to which the perpendicular edge is considered.  Thus, $N^{1 a}$ is the force density along the edge $2$. Likewise $Q^{2} \equiv N^{2 3}$ is the force density perpendicular to the shell in the $z$-direction along edge  $1$.

In \refeq{eq:nablaStressForm} the thickness integration  produces precisely such terms:  
\bea
\label{eq:Resultants1}
N^{a} = N^{a a} \continue
N^{1 a} = \int_{-h/2}^{h/2} \rmd z \, \sigma^{1 a} \, (1+z/R_{2}) \continue
N^{2 a} = \int_{-h/2}^{h/2} \rmd z \, \sigma^{2 a} \, (1+z/R_{1}) \continue
Q^{1} = \int_{-h/2}^{h/2} \rmd z \, \sigma^{1 3} \, (1+z/R_{2}) \continue
Q^{2} = \int_{-h/2}^{h/2} \rmd z \, \sigma^{2 3} \, (1+z/R_{1}) \, \,.
\eea
referred to as {\it stress resultants}. 
Also  corresponding moments are defined as:
\bea
\label{eq:Resultants2}
M^{a} = M^{a a} \continue
M^{1 a} = \int_{-h/2}^{h/2} \rmd z \, \sigma^{1 a} \, z \, (1+z/R_{2}) \continue
M^{2 a} = \int_{-h/2}^{h/2} \rmd z \, \sigma^{2 a} \, z \, (1+z/R_{1}) \, 
\eea 
following the conventions of  \cite{kraus}.

The resultants satisfy the constraint
\beq
\label{eq:resultantConstr} 
N^{1 2} -N^{2 1} = \frac{1}{R^2} \, M^{2 1} - \frac{1}{R_1} \, M^{1 2} \,.
\eeq
\refeq{eq:resultantConstr} follows from the symmetry of the stress tensor
\beq \fl
0 = \int \, \rmd z \, (\sigma^{1 2} - \sigma^{2 1}) (1+z/R_1)(1+ z/R_2) = N^{1 2 }+ \frac{1}{R_1} \, M^{1 2} - N^{2 1}- \frac{1}{R_2} \, M^{2 1}  \,.
\eeq

We express the resultants in tensor form as
\bea
N^{a b} &= \int _{-h/2}^{h/2} \rmd z \,  \, (\delta^a_c+z \, \dmod^{a}_c)\,  \sigma^{c b} \continue
Q^{a } &= \int _{-h/2}^{h/2} \rmd z \,  \, (\delta^a_c +z \,\dmod^{a}_c) \, \sigma^{c 3} \continue
M^{a b} &= \int _{-h/2}^{h/2} \rmd z \,  z \, (\delta^a_c + z \,\dmod^{a}_c) \,  \sigma^{c b}   
\eea
where a modified curvature tensor $\mathbf{\dmod}$ is introduced as
\beq \label{eq:dmod}
\mathbf{\dmod}= \frac{1}{R_2} \, \midbas{1} \otimes \midbas{1}  +\frac{1}{R_1} \, \midbas{2} \otimes \midbas{2}  = (\tr \mathbf{d}) \, \mathbf{a} - \mathbf{d} \,.
\eeq
For other  resultants, see \cite{fluegge}.

\subsubsection{Force balance for shell element}
We continue the example from \refsect{sec:covDerStressForm}.
When integrating over $z$ and multiplying with $A_{1} A_{2}$ note that $A_{a}$ and $R_{a}$ are independent of $z$. Also the symmetry $\sigma_{2 1} = \sigma_{1 2}$ is used to get a term $N^{1 2}$.  The integration over the $\partial_{3} (= \partial_z)$ -term is evaluated at the boundary.
Collecting terms  proportional to  $d\alpha^1 \wedge d\alpha^2$ gives the first equation of motion:
\bea
\label{eq:Momentum1}
\fl \partial_{1}(A_{2} N^{1 }) + \partial_{2}(A_{1} N^{2 1}) +A_{1,2} N^{1 2}- A_{2,1} N^{2} + A_{1} A_{2} (Q^{1}/R_{1}+q_{1}) = A_{1} A_{2} \rho h \ddot{u}^{1}
\eea
with the effective load 
\beq
\fl q^{1} = \left[ \sigma^{1 3} (1+z/R_{1})(1+z/R_{2}) \right]_{z=-h/2}^{h/2} + \int_{-h/2}^{h/2} \rmd z X^{1}\, (1+z/R_{1})(1+z/R_{2})  \, .
\eeq
Similar calculations are done for the second component $a=2$ in \refeq{eq:nablaStressForm}. 

\subsection{Moment conservation}
The moment in the direction ${\bf z} \wedge {\bf e}_{i}$ is found by multipliying   \refeq{microEquil}  with $z$. As before we consider  $i=a=1$ and 
the derivation is very similar to the above and differs only in a partial integration of the corresponding $\partial_{3}$-term
\bea
\label{eq:dzTerm}
\fl \int_{-h/2}^{h/2} \rmd z \,z \, \partial_{3}(\sigma^{3 1}(1+z/R_{1})(1+z/R_{2}))     \,   A_{1} A_{2} \,d\alpha^{1} \wedge d\alpha^{2} \\ \nonumber
\fl = \left( \left[ z \, \sigma^{3 1}(1+z/R_{1})(1+z/R_{2})  \right]_{z=-h/2}^{h/2} - \int_{-h/2}^{h/2} \rmd z  \, \sigma^{3 1}(1+z/R_{2}) \right)     \,   A_{1} A_{2} \,d\alpha^{1} \wedge d\alpha^{2} \\ \nonumber
\fl -  \int_{-h/2}^{h/2} \rmd z  \, \sigma^{3 1}(1+z/R_{2}) \, z/R_{1}     \,   A_{1} A_{2} \,d\alpha^{1} \wedge d\alpha^{2}
\eea
The last term cancels with 
\[
\int_{z = -h/2}^{h/2}   \, z \, \oneform{1}{3} \wedge \stressForm{3} =\int_{z = -h/2}^{h/2}   \, z \, \oneform{1}{3} \wedge \sigma^{31} \, \bas{2} \wedge dz \, .
\]
The second term in the parenthesis in \refeq{eq:dzTerm} gives the integral for $Q^{1}$, whereas the boundary term is grouped together with body moments as done above for the effective load $q^{i}$. 

\subsection{Normal resultants}
\label{sec:NormalResultant}
The final stress resultant involves the normal stress and is denoted $Q^a$. The calculation can be done as in  \refeq{eq:Momentum1} and stated   with derivatives of products of  Lam\'{e}-coefficients $A_i$ and resultants.  

However, the derivatives of the products can be expanded and written as covariant derivatives of  the mid-section using \refeq{eq:connectionMid}.   This can also be seen at an earlier stage in the calculation by  keeping the mid-section frame $\{  \midbas{1}, \midbas{2}, dz \}$  and evaluating derivatives using  \refeq{eq:NoTorsionMid}. For simplicity we show this alternative approach for $Q^{a}$ but similar calculations can be done for the other resultants.

The  covariant derivative becomes
\beq
\fl \bnabla \stressForm{3} = d  \stressForm{3} + \oneform{3}{a} \wedge \stressForm{a}   
\eeq
with
\bea
\fl  d  \stressForm{3} = d(\sigma^{1 3} \bas{2} \wedge dz)  + d(\sigma^{2 3} dz\wedge \bas{1}  ) + d(\sigma^{33} \bas{1} \wedge \bas{2})\continue
 \fl = (\partial_{\hat{1}} (\sigma^{1 3} (1+ z/R_2))+ \partial_{\hat{2}} (\sigma^{2 3} (1+ z/R_1)) +\partial_{\hat{3}} (\sigma^{3 3} (1+z/R_1)(1+ z/R_2)) ) \midbas{1}\wedge \midbas{2} \wedge dz  \continue
\fl - (\sigma^{1 3} (1+ z/R_2)) \midoneform{2}{1} \wedge \midbas{1} \wedge dz + (\sigma^{2 3} (1+ z/R_1)) dz \wedge \midoneform{1}{2} \wedge \midbas{2}  \continue
\fl =\left[ \partial_{\hat{1}} (\sigma^{1 3} (1+ z/R_2))+ \partial_{\hat{2}} (\sigma^{2 3} (1+ z/R_1)) +\partial_{\hat{3}} (\sigma^{3 3} (1+z/R_1)(1+ z/R_2))    \right.  \continue
\left. \fl +   (\sigma^{1 3} (1+ z/R_2)) \, \widetilde{\Gamma}^{2}_{2 1} +  (\sigma^{2 3} (1+ z/R_1)) \, \widetilde{\Gamma}^{1}_{1 2}  \right] \midbas{1} \wedge \midbas{2} \wedge dz   
\eea
and
\beq
\fl \oneform{3}{a} \wedge \stressForm{a} 
 =  -\frac{\bas{a}}{R_a + z} \wedge \sigma^{b a}  \hdual{b}  =   - \frac{1}{R_a + z}   \sigma^{b a}  \delta_{a b}  \, \volel  
\eeq
so
\bea
\fl \int    \bnabla \stressForm{3}  
= \left( \partial_{\hat{1}} Q^1 + \widetilde{\Gamma}^{1}_{1 2} Q^2+ \partial_{\hat{2}} Q^2 + \widetilde{\Gamma}^{2}_{2 1} Q^1  - d_{a b} N^{a b} \right. \continue
\fl  \left. +  \left[ \sigma^{3 3} (1+z/R_{1})(1+z/R_{2}) \right]_{z=-h/2}^{h/2}   \right)  \midbas{1} \wedge \midbas{2} 
 \equiv \left[ \widetilde{\nabla}_a Q^a - d_{a b} N^{a b} +q^3 \right]  \midbas{1} \wedge \midbas{2}  \,.
\eea
as  $\widetilde{\Gamma}^{i}_{j i} \equiv \iota_{\widetilde{\mathbf{e}_j}} \midoneform{i}{i}= 0$ for all $i,j$  (no summation) by the antisymmetry of the frame connection coefficients.
On the other hand, the kinematic body force equals
\beq
\int_{z=-h/2}^{h/2} \, \rho \ddot{u}^3 \volel  \equiv  \rho h \ddot{w} \,  \midbas{1} \wedge \midbas{2} +O(h^3)\,. 
\eeq
Thus to leading order
\beq
\widetilde{\nabla}_a Q^a - d_{a b} N^{a b} +q^3 = \rho h \ddot{w} \, .
\eeq

\subsection{Dynamical equations of motion} 
In summary, using exterior calculus and moving frames one finds directly  the  shell equations  in classical form  \cite{kraus}:
\bea
\label{EOM}
\fl \partial_{1}(A_{2} N^{1 }) + \partial_{2}(A_{1} N^{2 1}) +A_{1,2} N^{1 2}- A_{2,1} N^{2} + A_{1} A_{2} (Q^{1}/R_{1}+q^{1}) &= A_{1} A_{2} \rho h \ddot{u}^{1} \continue 
\fl \partial_{1}(A_{2} N^{1 2}) + \partial_{2}(A_{1} N^{2 }) +A_{2, 1} N^{2 1}- A_{1,2} N^{1} + A_{1} A_{2} (Q^{2}/R_{2}+q^{2}) &= A_{1} A_{2} \rho h \ddot{u}^{2} \continue
\fl \partial_{1}(A_{2} Q^{1 }) + \partial_{2}(A_{1} Q^{2 }) - A_{1} A_{2} ({N^{1}}/{R_{1}} +{N^{2}}/{R_{2}}) + A_{1} A_{2} q^{3} &= A_{1} A_{2} \rho h \ddot{w} \continue
\fl \partial_{1}(A_{2} M^{1 }) + \partial_{2}(A_{1} M^{2 1}) +A_{1,2} M^{1 2}- A_{2,1} M^{2} - A_{1} A_{2} (Q^{1}-m^{1}) &= 0 \continue
\fl \partial_{1}(A_{2} M^{1 2}) + \partial_{2}(A_{1} M^{2 }) +A_{2, 1} M^{2 1}- A_{1,2} M^{1} -A_{1} A_{2} (Q^{2} -m^{2}) &= 0 \, .
\eea 
The right hand sides, the {\it kinematic} part,  are also quantities integrated over thickness. They will be discussed in \refsect{sec:acc}.  

Yet, the equations for the resultants can also be written {\it covariantly}:
\bea
\widetilde{\nabla}_b N^{b a} + d^a_b \, Q^b  + q^a &= \rho h \ddot{u}^a \continue
\widetilde{\nabla}_a Q^a - N^{a b} \, d_{b a} + q^3 &= \rho h \ddot{w} \continue
\widetilde{\nabla}_b M^{b a} -Q^a +m^a&=   0 \,,
\eea
since in \refeq{EOM} in the equations for $N^{a b}$ and $M^{a b}$  all the first four terms come from a tangential covariant derivative when we recognize  the  connection coefficients given in  \refeq{eq:connectionMid}. Likewise for $Q^a$ as demonstrated in \refsect{sec:NormalResultant}.

Eliminating the normal stress $Q^a$ and assuming no body forces and moments gives
\bea
\widetilde{\nabla}_b N^{b a} + d^a_b \,   \widetilde{\nabla}_c \,  M^{c b}   &= \rho h \ddot{u}^a \continue
\widetilde{\nabla}_a  \widetilde{\nabla}_b M^{b a}   - N^{a b} \, d_{b a}  &= \rho h \ddot{w} \,.
\eea

\subsection{Inertia terms}
\label{sec:acc}
When the acceleration displacement field $\mathbf{\ddot{u}}$ is integrated over thickness only those terms constant in $z$ survive leading to the right hand sides of the three first equations in  \refeq{EOM}.
  But for the last two, the moment equations,  \refeq{microEquil} is already multiplied with $z$ and only the acceleration of $\beta^{i}$  remains.  This leads to terms 
\[
 {h^{3} \over 12 } \, A_{1} A_{2} \, \rho  \, \ddot{\beta}^{i} \, ,
\]
which go as the cube of thickness and are neglected in this treatment.

\subsection{Plane stress}
Although all that  we wanted to consider concerning moving frames and shells was presented in the previous sections we mention for completeness one final subject, which is  how  to relate the resultants with the displacements.  Here a link between stress and strain is needed.  These are in the form of {\it  constitutive equations} and can be thought of as a generalization of Hooke's law.

In the analysis of plates and shell the so-called plane stress approximation is used \cite{kraus,fluegge}. In that approximation, the stress-tensor  has almost the same form as in bulk  elasticity  except that   the coefficients of elasticity  are slightly altered. We refer the reader to \cite{kraus,fluegge} for a full discussion  and just state the approximation in the case of isotropic materials
\beq
\label{eq:planeStress}
\sigma^{a b} = \frac{E \nu}{1-\nu^{2}} \, a^{a b}\,  \tr \bepsilon + \frac{E}{1+\nu} \, \epsilon^{a b}  \,.
\eeq
Following the notation of \cite{Norris94} a 4-tensor is introduced
\beq
H^{a b c d} = \frac{1-\nu}{2} \left(a^{a c} a^{b d} + a^{a d} a^{b c} \right) + \nu \, a^{a b} \, a^{c d}
\eeq
such that
\beq
\sigma^{a b} = \frac{E}{1-\nu^2} \,  H^{a b c d} \, \epsilon_{c d} \,.
\eeq

\subsubsection{Stress resultants}
We are now prepared to  integrate \refeq{eq:Resultants1} to get the resultants. Separate the stress tensor according to the different orders in  the expansion in $z$:
\beq
\label{eq:planeStressOrder}
\sigma^{a b (i)} = \frac{E }{1-\nu^2} \, H^{a b c d} \, \epsilon^{(i)}_{c d} \,
\eeq
for $i=0,1$ with $\bepsilon^{(i)}$ given by \refeq{eq:strainExpansion}. Plain calculation from \refeq{eq:Resultants1} and \refeq{eq:planeStress} gives the resultants as
\bea \label{eq:resultantExpansion}
N^{a b} &= h \, \sigma^{a b (0)} +\frac{h^3}{12}\, {\dmod}^{a}_c \, \sigma^{c b (1)} \continue
M^{a b} &= \frac{h^3}{12}\, \left(  \sigma^{a b (1)}+{\dmod}^{a}_c \, \sigma^{c b (0)}\right) 
\eea
using $\mathbf{\dmod}$ defined by \refeq{eq:dmod}.
Or, in terms of the plane stress tensor $H^{a b c d}$
\bea \label{eq:resultantExpansion1}
N^{a b} &= C \, H^{a b c f} \, \epsilon^{(0)}_{c f}  +B \,{\dmod}^{a}_c \,  H^{c b f e } \, \epsilon^{(1)}_{f e} \continue
M^{a b} &= B \, \left( H^{a b c f} \,  \epsilon^{(1)}_{c f }+{\dmod}^{a}_c \, H^{c b f e} \, \epsilon^{(0)}_{f e} \,.\right)
\eea
The constants $C$ and $B$ are the stretching and bending rigidities
\beq
C = \frac{E h }{1-\nu^2} \qquad \mbox{with} \qquad B= \frac{E h^3 }{12(1-\nu^2)} \,.
\eeq

In \refeq{eq:resultantExpansion} the factors $z/R_{i}$ from  \refeq{eq:Resultants1} are included leading to corrections proportional to $h^3/R_i$, see \cite{kraus, fluegge}.  If included the relation \refeq{eq:resultantConstr} among the resultants is satisfied. Thus, some tensor algebra on \refeq{eq:resultantExpansion}  shows
\beq \label{eq:Equil6}
\varepsilon_{a b} N^{a b} = -\varepsilon_{a b} d^a_c M^{c b} \, .
\eeq
Therefore  the antisymmetric part of  $\mathbf{ N}$ equals that of  $\mathbf{d \cdot M}$ and in lines of curvature coordinates  \refeq{eq:Equil6} has the form
\beq
N^{1 2}-N^{2 1} = \frac{1}{R_2} \, M^{2 1} - \frac{1}{R_1} \, M^{1 2}
\eeq
which coincides with \refeq{eq:resultantConstr}.

\section{Summary and conclusion}
The method of  moving frames and exterior calculus allow a fast derivation of the equations of motion for a curved elastic shell.  First,  the kinematic stretching and bending field  corresponds to a particular  strain tensor field obtained using the Lie-derivative  on the metric.  Second, the equations describing the dynamics are found using   stress forms which are differential two-forms encoding the stress tensor in a convenient way.  Finally, all equations are seen to be covariant and hence valid in any coordinate system.

\ack N.S. thanks the Swedish Research Council.

%\tableofcontents

\indent 
\end{document}